\documentclass[11pt]{article}
\pdfoutput=1
\usepackage{graphicx}
\usepackage{amssymb}
\usepackage{amsmath}
\usepackage[left]{lineno}
\usepackage{subfigure}
\usepackage{booktabs}
\usepackage{multirow}
\usepackage[title]{appendix}
\usepackage{lineno}

\newcommand{\pp}      {$K^+\rightarrow\pi^+\pi^0$~}
\newcommand{\ppg}     {$K^+\rightarrow\pi^+\pi^0(\gamma)$~}

\newcommand{\pppc}    {$K^+\rightarrow\pi^+\pi^+\pi^-$~}
\newcommand{\kethree} {$K^+\rightarrow\pi^0e^+\nu$}
\newcommand{\kefour}  {$K^+\rightarrow \pi^+\pi^- e^+\nu$~}
\newcommand{\mn}      {$K^+\rightarrow\mu^+\nu$~}
\newcommand{\mng}     {$K^+\rightarrow\mu^+\nu(\gamma)$~}
\newcommand{\kmthree} {$K^+\rightarrow\mu^+\pi^0\nu$}
\newcommand{\ppgr}    {$K^+\rightarrow\pi^+\pi^0\gamma$~}
\newcommand{\pnnc}    {$K^+\rightarrow\pi^+\nu\bar{\nu}$~}
\newcommand{\pnnz}    {$K_L\rightarrow\pi^0\nu\bar{\nu}$~}

\newcommand{\pic}     {$\pi^+$~}
\newcommand{\mup}     {$\mu^+$~}
\newcommand{\kp}      {$K^+$~}
\newcommand{\ppic}    {$P_{\pi^+}$~}
\def\mmis{m^2_\mathrm{miss}}

\begin{document}
\pagenumbering{arabic}
\centerline{\LARGE EUROPEAN ORGANIZATION FOR NUCLEAR RESEARCH}

\begin{flushright}
CERN-EP-2018-314  \\
November 15, 2018 \\
 \vspace{2mm}
Revised version\\
December 21, 2018\\\end{flushright}
\vspace{15mm}

\begin{center}
\Large{\bf First search for {\boldmath\pnnc}using the decay-in-flight technique \\
\vspace{5mm}
}
The NA62 Collaboration
\end{center}
\vspace{10mm}

\begin{abstract}
The NA62 experiment at the CERN SPS reports the first search for $K^+ \rightarrow \pi^+ \nu \bar{\nu}$ using the decay-in-flight technique, based on a sample of 
$1.21\times10^{11}$ \kp~decays collected in 2016. The single event sensitivity is  $3.15\times 10^{-10}$, corresponding to 0.267 Standard Model events.   One signal candidate is observed while  the 
 expected background is 0.152 events. 
 This leads to an upper limit of $14 \times 10^{-10}$ on the $K^+ \rightarrow \pi^+ \nu \bar{\nu}$  
branching ratio at 95\% CL.
\end{abstract}
\vspace{10mm}
\begin{center}
This paper is dedicated to the memory of our colleagues S. Balev and F. Hahn.
\end{center}
\vspace{5mm}
\begin{center}
\em{Accepted for publication in Physics Letters B}
\end{center}
\clearpage
\begin{center}
{\Large The NA62 Collaboration$\,$\renewcommand{\thefootnote}{\fnsymbol{footnote}}%
\footnotemark[1]\renewcommand{\thefootnote}{\arabic{footnote}}}\\
\end{center}
\vspace{3mm}
\begin{raggedright}
\noindent
{\bf Universit\'e Catholique de Louvain, Louvain-La-Neuve, Belgium}\\
 E.~Cortina Gil,
 E.~Minucci,
 S.~Padolski$\,$\footnotemark[1],
 P.~Petrov,
 B.~Velghe$\,$\footnotemark[2]\\[2mm]

{\bf Faculty of Physics, University of Sofia, Sofia, Bulgaria}\\
 G.~Georgiev$\,$\footnotemark[3],
 V.~Kozhuharov$\,$\footnotemark[3],
 L.~Litov\\[2mm]

{\bf TRIUMF, Vancouver, British Columbia, Canada}\\
 T.~Numao\\[2mm]

{\bf University of British Columbia, Vancouver, British Columbia, Canada}\\
 D.~Bryman,
 J.~Fu$\,$\footnotemark[4]\\[2mm]

{\bf Charles University, Prague, Czech Republic}\\
 T.~Husek$\,$\footnotemark[5],
 K.~Kampf,
 M.~Zamkovsky\\[2mm]

{\bf Institut f\"ur Physik and PRISMA Cluster of excellence, Universit\"at Mainz, Mainz, Germany}\\
 R.~Aliberti,
  G.~Khoriauli$\,$\footnotemark[6],
 J.~Kunze,
 D.~Lomidze$\,$\footnotemark[7],
 R.~Marchevski$\,$\renewcommand{\thefootnote}{\fnsymbol{footnote}}%
\footnotemark[1]$^,$\renewcommand{\thefootnote}{\arabic{footnote}}%
\footnotemark[8],
 L.~Peruzzo,
 M.~Vormstein,
 R.~Wanke\\[2mm]

{\bf Dipartimento di Fisica e Scienze della Terra dell'Universit\`a e INFN, Sezione di Ferrara, Ferrara, Italy}\\
 P.~Dalpiaz,
 M.~Fiorini,
 E.~Gamberini$\,$\footnotemark[8],
 I.~Neri,
 A.~Norton,
 F.~Petrucci,
 H.~Wahl\\[2mm]

{\bf INFN, Sezione di Ferrara, Ferrara, Italy}\\
 A.~Cotta Ramusino,
 A.~Gianoli\\[2mm]

{\bf Dipartimento di Fisica e Astronomia dell'Universit\`a e INFN, Sezione di Firenze, Sesto Fiorentino, Italy}\\
 E.~Iacopini,
 G.~Latino,
 M.~Lenti\\[2mm]

{\bf INFN, Sezione di Firenze, Sesto Fiorentino, Italy}\\
 A.~Bizzeti$\,$\footnotemark[9],
 F.~Bucci,
 R.~Volpe$\,$\footnotemark[10]\\[2mm]

{\bf Laboratori Nazionali di Frascati, Frascati, Italy}\\
 A.~Antonelli,
 G.~Lamanna$\,$\footnotemark[11],
 G.~Lanfranchi,
 G.~Mannocchi,
 S.~Martellotti,
 M.~Moulson,
 M.~Raggi$\,$\footnotemark[12],
 T.~Spadaro\\[2mm]

{\bf Dipartimento di Fisica ``Ettore Pancini'' e INFN, Sezione di Napoli, Napoli, Italy}\\
 F.~Ambrosino,
 T.~Capussela,
 M.~Corvino,
 D.~Di Filippo,
 P.~Massarotti,
 M.~Mirra,
 M.~Napolitano,
 G.~Saracino\\[2mm]

{\bf Dipartimento di Fisica e Geologia dell'Universit\`a e INFN, Sezione di Perugia, Perugia, Italy}\\
 G.~Anzivino,
 F.~Brizioli,
 E.~Imbergamo,
 R.~Lollini,
 C.~Santoni\\[2mm]

{\bf INFN, Sezione di Perugia, Perugia, Italy}\\
 M.~Barbanera$\,$\footnotemark[13],
 P.~Cenci,
 B.~Checcucci,
 V.~Duk$\,$\footnotemark[14],
 P.~Lubrano,
 M.~Lupi$\,$\footnotemark[8],
 M.~Pepe,
 M.~Piccini\\[2mm]

{\bf Dipartimento di Fisica dell'Universit\`a e INFN, Sezione di Pisa, Pisa, Italy}\\
 F.~Costantini,
 L.~Di Lella,
 N.~Doble,
 M.~Giorgi,
 S.~Giudici,
 E.~Pedreschi,
 M.~Sozzi\\[2mm]

{\bf INFN, Sezione di Pisa, Pisa, Italy}\\
 C.~Cerri,
 R.~Fantechi,
 R.~Piandani$\,$\footnotemark[15],
 J.~Pinzino$\,$\footnotemark[8],
 L.~Pontisso,
 F.~Spinella\\[2mm]

{\bf Scuola Normale Superiore e INFN, Sezione di Pisa, Pisa, Italy}\\
 I.~Mannelli\\[2mm]

{\bf Dipartimento di Fisica, Sapienza Universit\`a di Roma e INFN, Sezione di Roma I, Roma, Italy}\\
 G.~D'Agostini\\[2mm]

{\bf INFN, Sezione di Roma I, Roma, Italy}\\
 A.~Biagioni,
 E.~Leonardi,
 A.~Lonardo,
 P.~Valente,
 P.~Vicini\\[2mm]

{\bf INFN, Sezione di Roma Tor Vergata, Roma, Italy}\\
 R.~Ammendola,
 V.~Bonaiuto$\,$\footnotemark[16],
 L.~Federici$\,$\footnotemark[8],
 A.~Fucci,
 A.~Salamon,
 F.~Sargeni$\,$\footnotemark[17]\\[2mm]

{\bf Dipartimento di Fisica dell'Universit\`a e INFN, Sezione di Torino, Torino, Italy}\\
 R.~Arcidiacono$\,$\footnotemark[18],
 B.~Bloch-Devaux,
 M.~Boretto$\,$\footnotemark[8],
 E.~Menichetti,
 E.~Migliore,
 D.~Soldi\\[2mm]

{\bf INFN, Sezione di Torino, Torino, Italy}\\
 C.~Biino,
 A.~Filippi,
 F.~Marchetto\\[2mm]

{\bf Instituto de F\'isica, Universidad Aut\'onoma de San Luis Potos\'i, San Luis Potos\'i, Mexico}\\
 J.~Engelfried,
 N.~Estrada-Tristan$\,$\footnotemark[19]\\[2mm]

{\bf Horia Hulubei national Institute of Physics and Nuclear Engineering, Bucharest-Magurele, Romania}\\
 A. M.~Bragadireanu,
 S. A.~Ghinescu,
 O. E.~Hutanu\\[2mm]

{\bf Joint Institute for Nuclear Research, Dubna, Russia}\\
 T.~Enik,
 V.~Falaleev,
 V.~Kekelidze,
 A.~Korotkova,
 D.~Madigozhin,
 M.~Misheva$\,$\footnotemark[20],
 N.~Molokanova,
 S.~Movchan,
 I.~Polenkevich,
 Yu.~Potrebenikov,
 S.~Shkarovskiy,
 A.~Zinchenko$\,$\renewcommand{\thefootnote}{\fnsymbol{footnote}}\footnotemark[2]\renewcommand{\thefootnote}{\arabic{footnote}}\\[2mm]

{\bf Institute for Nuclear Research of the Russian Academy of Sciences, Moscow, Russia}\\
 S.~Fedotov,
 E.~Gushchin,
 A.~Khotyantsev,
 A.~Kleimenova$\,$\footnotemark[10],
 Y.~Kudenko$\,$\footnotemark[21],
 V.~Kurochka,
 M.~Medvedeva,
 A.~Mefodev,
 A.~Shaikhiev$\,$\footnotemark[10]\\[2mm]

{\bf Institute for High Energy Physics - State Research Center of Russian Federation, Protvino, Russia}\\
 S.~Kholodenko,
 V.~Kurshetsov,
 V.~Obraztsov,
 A.~Ostankov,
 V.~Semenov,
 V.~Sugonyaev,
 O.~Yushchenko\\[2mm]

{\bf Faculty of Mathematics, Physics and Informatics, Comenius University, Bratislava, Slovakia}\\
 L.~Bician,
 T.~Blazek,
 V.~Cerny,
 M.~Koval$\,$\footnotemark[8],
 Z.~Kucerova\\[2mm]

{\bf CERN,  European Organization for Nuclear Research, Geneva, Switzerland}\\
 A.~Ceccucci,
 H.~Danielsson,
 N.~De Simone$\,$\footnotemark[22],
 F.~Duval,
 B.~D\"obrich,
 L.~Gatignon,
 R.~Guida,
 F.~Hahn$\,$\renewcommand{\thefootnote}{\fnsymbol{footnote}}\footnotemark[2]\renewcommand{\thefootnote}{\arabic{footnote}},
 B.~Jenninger,
 P.~Laycock$\,$\footnotemark[1],
 G.~Lehmann Miotto,
 P.~Lichard,
 A.~Mapelli,
 K.~Massri,
 M.~Noy,
 V.~Palladino$\,$\footnotemark[23],
 M.~Perrin-Terrin$\,$\footnotemark[24]$^,\,$\footnotemark[25],
 V.~Ryjov,
 S.~Venditti\\[2mm]

{\bf University of Birmingham, Birmingham, United Kingdom}\\
 M. B.~Brunetti,
 V.~Fascianelli$\,$\footnotemark[26],
 F.~Gonnella,
 E.~Goudzovski,
 L.~Iacobuzio,
 C.~Lazzeroni,
 N.~Lurkin,
 F.~Newson,
 C.~Parkinson,
 A.~Romano,
 A.~Sergi,
 A.~Sturgess,
 J.~Swallow\\[2mm]

{\bf University of Bristol, Bristol, United Kingdom}\\
 H.~Heath,
 R.~Page,
 S.~Trilov\\[2mm]

{\bf University of Glasgow, Glasgow, United Kingdom}\\
 B.~Angelucci,
 D.~Britton,
 C.~Graham,
 D.~Protopopescu\\[2mm]

{\bf University of Liverpool, Liverpool, United Kingdom}\\
 J. B.~Dainton$\,$\footnotemark[27],
 J. R.~Fry$\,$\footnotemark[14],
 L.~Fulton,
 D.~Hutchcroft,
 E.~Maurice$\,$\footnotemark[28],
 G.~Ruggiero$\,$\renewcommand{\thefootnote}{\fnsymbol{footnote}}%
\footnotemark[1]$^,$\renewcommand{\thefootnote}{\arabic{footnote}}%
\footnotemark[27],
 B.~Wrona\\[2mm]

{\bf George Mason University, Fairfax, Virginia, USA}\\
 A.~Conovaloff,
 P.~Cooper,
 D.~Coward$\,$\footnotemark[29],
 P.~Rubin\\[2mm]

\end{raggedright}
%
%
\setcounter{footnote}{0}
\renewcommand{\thefootnote}{\fnsymbol{footnote}}
\footnotetext[1]{Corresponding authors: G.~Ruggiero, R.~Marchevski,\\ email:giuseppe.ruggiero@cern.ch, radoslav.marchevski@cern.ch}
\footnotetext[2]{Deceased}
\renewcommand{\thefootnote}{\arabic{footnote}}
\footnotetext[1]{Present address: Brookhaven National Laboratory, Upton, NY 11973, USA}
\footnotetext[2]{Present address: TRIUMF, Vancouver, British Columbia, V6T 2A3, Canada}
\footnotetext[3]{Also at Laboratori Nazionali di Frascati, I-00044 Frascati, Italy}
\footnotetext[4]{Present address: UCLA Physics and Biology in Medicine, Los Angeles, CA 90095, USA}
\footnotetext[5]{Present address: IFIC, Universitat de Val\`encia - CSIC, E-46071 Val\`encia, Spain}
\footnotetext[6]{Present address: Universit\"at W\"urzburg, D-97070 W\"urzburg, Germany}
\footnotetext[7]{Present address: Universit\"at Hamburg, D-20146 Hamburg, Germany}
\footnotetext[8]{Present address: CERN,  European Organization for Nuclear Research, CH-1211 Geneva 23, Switzerland}
\footnotetext[9]{Also at Dipartimento di Fisica, Universit\`a di Modena e Reggio Emilia, I-41125 Modena, Italy}
\footnotetext[10]{Present address: Universit\'e Catholique de Louvain, B-1348 Louvain-La-Neuve, Belgium}
\footnotetext[11]{Present address: Dipartimento di Fisica dell'Universit\`a e INFN, Sezione di Pisa, I-56100 Pisa, Italy}
\footnotetext[12]{Present address: Dipartimento di Fisica, Universit\`a di Roma La Sapienza, I-00185 Roma, Italy}
\footnotetext[13]{Present address: INFN, Sezione di Pisa, I-56100 Pisa, Italy}
\footnotetext[14]{Present address: School of Physics and Astronomy, University of Birmingham, Birmingham, B15 2TT, UK}
\footnotetext[15]{Present address: Dipartimento di Fisica e Geologia dell'Universit\`a e INFN, Sezione di Perugia, I-06100 Perugia, Italy}
\footnotetext[16]{Also at Department of Industrial Engineering, University of Roma Tor Vergata, I-00173 Roma, Italy}
\footnotetext[17]{Also at Department of Electronic Engineering, University of Roma Tor Vergata, I-00173 Roma, Italy}
\footnotetext[18]{Also at Universit\`a degli Studi del Piemonte Orientale, I-13100 Vercelli, Italy}
\footnotetext[19]{Also at Universidad de Guanajuato, Guanajuato, Mexico}
\footnotetext[20]{Present address: Institute of Nuclear Research and Nuclear Energy of Bulgarian Academy of Science (INRNE-BAS), BG-1784 Sofia, Bulgaria}
\footnotetext[21]{Also at National Research Nuclear University (MEPhI), 115409 Moscow and Moscow Institute of Physics and Technology, 141701 Moscow region, Moscow, Russia}
\footnotetext[22]{Present address: DESY, D-15738 Zeuthen, Germany}
\footnotetext[23]{Present address: Physics Department, Imperial College London, London, SW7 2BW, UK}
\footnotetext[24]{Present address: Centre de Physique des Particules de Marseille, Universit\'e Aix Marseille, CNRS/IN2P3, F-13288, Marseille, France}
\footnotetext[25]{Also at Universit\'e Catholique de Louvain, B-1348 Louvain-La-Neuve, Belgium}
\footnotetext[26]{Present address: Dipartimento di Psicologia, Universit\`a di Roma La Sapienza, I-00185 Roma, Italy}
\footnotetext[27]{Present address: Physics Department, University of Lancaster, Lancaster, LA1 4YW, UK}
\footnotetext[28]{Present address: Laboratoire Leprince Ringuet, F-91120 Palaiseau, France}
\footnotetext[29]{Also at SLAC National Accelerator Laboratory, Stanford University, Menlo Park, CA 94025, USA}

\clearpage
\section{Introduction}
\label{sec:intro}
The  flavour-changing neutral current decay  \pnnc    proceeds at the lowest order in the Standard Model (SM)
through electroweak box and penguin diagrams largely dominated by t-quark exchange.  The quadratic 
GIM mechanism and the small value of the CKM element $|V_\mathrm{td}|$ describing the transition of a top  into a down quark make this process extremely rare. Using  tree-level elements of 
the CKM matrix as external inputs, the SM predicts \cite{pnntheo1} the branching ratio to be ${\rm BR} = (8.4\pm1.0)\times10^{-11}$, where the uncertainty is dominated by the current precision on 
the CKM parameters. The intrinsic theoretical accuracy is at the  2\% level, as the computation includes NLO (NNLO) QCD corrections 
to the top (charm) quark contribution~\cite{pnntheo2,pnntheo3} and NLO electroweak corrections~\cite{pnntheo4}. Moreover, the hadronic matrix element 
largely cancels when normalised to the precisely measured BR of the $K^+\rightarrow\pi^0e^+\nu$ decay, with isospin breaking and non-perturbative 
effects calculated in detail~\cite{pnntheo4,pnntheo5}.

The \pnnc decay is sensitive to physics beyond the SM.  
The largest deviations from the SM are expected in models with new sources of flavour violation, where constraints from $B$ 
physics are weaker~\cite{pnnlh,pnnrs}. The experimental value 
of the CP-violation parameter $\varepsilon_K$ limits the expected $\text{BR}(K^+\rightarrow\pi^+\nu\bar{\nu})$ range within models with currents of defined chirality, 
producing specific correlation patterns between the \pnnc and \pnnz decay rates~\cite{pnnzz}. 
Present experimental constraints limit the range of variation within supersymmetric models~\cite{pnnsusy1,pnnsusy2,pnnsusy3}.
The \pnnc decay can also be sensitive to effects of lepton flavour non-universality~\cite{pnnlu}  and can constrain leptoquark 
models~\cite{pnnlq} aiming to explain the measured value of the CP-violation parameter  $\varepsilon'/\varepsilon$~\cite{pdg}.

The E787 and E949 experiments at BNL~\cite{bnl1,bnl2} studied the \pnnc decay using a decay-at-rest technique and
obtained $\text{BR}=(17.3^{+11.5}_{-10.5})\times10^{-11}$. The NA62 experiment at the CERN SPS aims to measure 
$\text{BR}(K^+\rightarrow\pi^+\nu\bar{\nu})$  more precisely   with a 
novel  decay-in-flight technique. This letter reports a result from the analysis of data collected by NA62 in 2016, corresponding to 2\% of the full statistics accumulated during the 2016--2018 data-taking period. 

\section{Beam line and detector}\label{sec:det}
The choice of the decay-in-flight technique is motivated by the possibility of obtaining an integrated flux of $\cal O$$(10^{13})$ kaon decays over a few years of data-taking with a signal acceptance of  a few percent, leading to the collection of $\cal  O$(100)~SM events in the $K^+ \to \pi^+ \nu \bar\nu$ channel. This technique utilises a high energy  
beam to boost  the kaon decay products to high energy where their detection and identification can be efficient.  The boost folds the decay products into a small angular region close to  the beam axis allowing for a classic fixed target geometry experiment.  
The detector should measure the incoming kaon and the outgoing pion while reducing 
background to the level of $10^{-11}$ of accepted kaon decays in the high rate environment necessary to achieve the required sensitivity.  
To this end, the  detector consists of a collection of sub-detectors, each designed to perform one main task, and  with complementary performance. 

\begin{figure}[t]
  \begin{center}
    \includegraphics[width=1.0\textwidth]{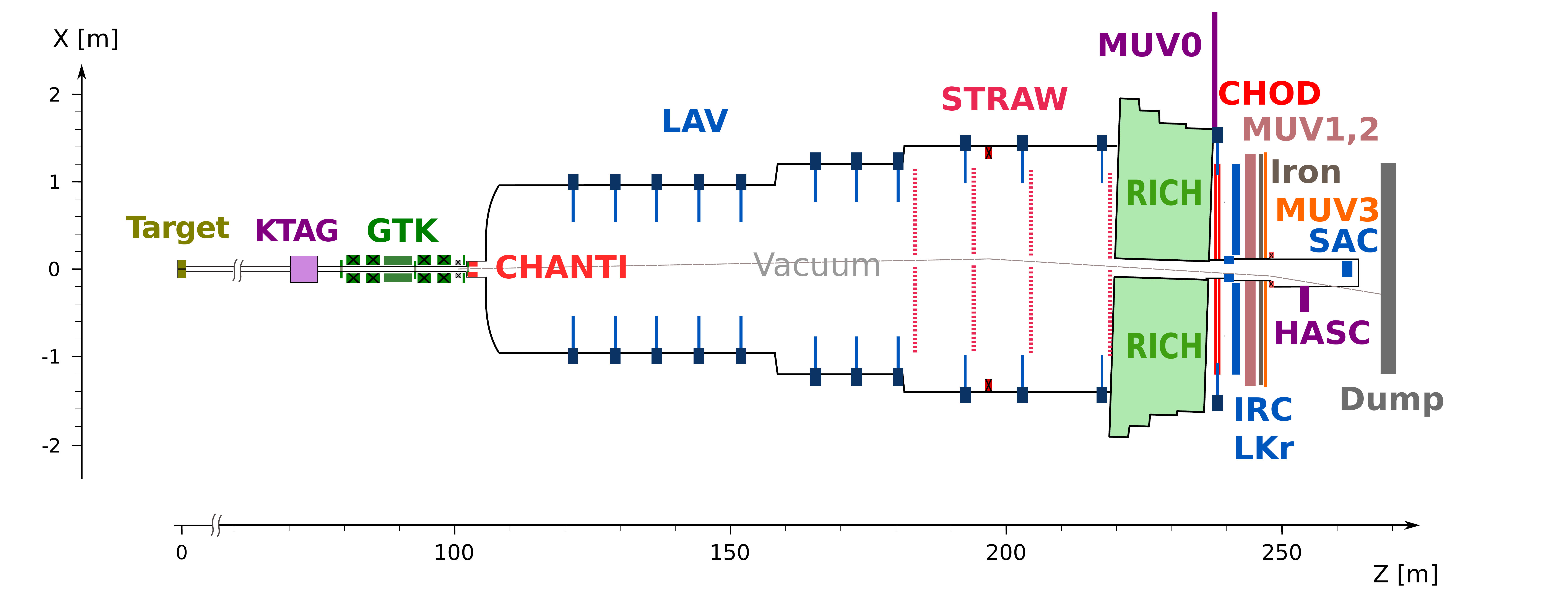}
    \caption{Schematic top view of the NA62 beam line and detector. Dipole magnets are displayed as boxes with superimposed crosses. The trajectory is shown of an un-decayed beam particle 
    in vacuum, crossing the detector apertures which avoid interactions with material. A dipole magnet between MUV3 and SAC deflects the charged particles of the beam out of the SAC acceptance.}
    \label{fig:layout}
  \end{center}
\end{figure}

The NA62 beam line and detector layout is  described in detail in~\cite{na62det} and shown in Fig.~\ref{fig:layout}    
together with the scale and reference system: 
the beam line defines the Z axis with its origin at the kaon production target and beam particles travelling in the positive direction, the Y axis points vertically up, and the X axis is horizontal and directed to form a right-handed coordinate system.

A secondary hadron beam of positive charge containing 70\%  $\pi^+$, 23\% protons and 6\% $K^+$, with a nominal momentum of 75~GeV/$c$ and 1\% rms momentum bite,  is derived from  400~GeV/$c$ protons extracted  from the SPS in spills of 3~s effective duration and interacting with a 40~cm long beryllium target. 
Typical intensities  for the present measurement range from  $1.0$ to $1.3\times10^{12}$ protons per pulse ($ppp$). 
 
 The time, momentum and direction of the charged components of the $K^+ \to \pi^+ \nu \bar\nu$  decay  are measured by the following detectors. A differential Cherenkov counter (KTAG, filled with N$_2$ at 1.75~bar pressure and read out by PMs grouped in eight sectors) tags incoming kaons with a 70 ps time resolution. Three silicon pixel stations (GTK) located before, between and after two pairs of dipole magnets (beam achromat), form a  spectrometer to measure momentum, direction and time of beam particles  with  0.15 GeV/$c$, 16 $\mu$rad and 100 ps resolutions. 
 A magnetic spectrometer (STRAW, comprising two pairs of straw chambers on either side of a dipole magnet)  measures the  momentum-vector of the outgoing particle with a momentum resolution  $\sigma_p / p$ in the  0.3--0.4\% range.  
 A ring-imaging Cherenkov counter (RICH, filled with neon  at atmospheric pressure) tags the decay particle with a precision better than 100~ps and provides particle identification.
 
 Two   scintillator hodoscopes  (CHOD, a matrix of tiles read out by SiPMs and NA48-CHOD, composed of two orthogonal planes of slabs, reused from the NA48 experiment) are used for triggering and timing purposes, providing  a 99\% efficient trigger and a time measurement with 200 ps resolution for charged particles.
 
The third GTK station (GTK3) is immediately preceded by a final collimator to partly block particles produced in upstream decays,
and marks the beginning of a 117~m-long vacuum tank. 
The first 80~m of the tank define a fiducial volume (FV) in which 13\% of the kaons decay. The beam has a  rectangular transverse profile of 52 $\times$ 24 mm$^2$ and a divergence of 0.11~mrad (rms) in each plane at the FV entrance. The typical beam particle  rate is 300~MHz. 

  Other  sub-detectors are used as  vetoes to suppress decays  into  photons or  multiple charged particles (electrons, pions or muons) or as complementary particle-identifiers.
Six stations of plastic scintillator bars (CHANTI) detect with 99\% efficiency and 1~ns time resolution  extra activity including inelastic interactions in  GTK3.
Twelve stations of ring-shaped electromagnetic calorimeters  (LAV1 to LAV12, made of lead-glass blocks) surround  the vacuum tank and the detector   to achieve hermetic acceptance for photons emitted in $K^+$ decays in the FV at polar angles between 10 and 50~mrad.
A 27 radiation length thick quasi-homogeneous liquid krypton electromagnetic calorimeter (LKr) detects photons from $K^+$  decays emitted at angles between 1 and 10~mrad and complements the RICH for particle identification. 
The LKr energy, spatial and time resolutions in NA62 conditions are $\sigma_E / E = 1.4\%$  at an energy deposit of 25~GeV,
1~mm and between 0.5 and 1~ns, respectively, depending on the amount and type of energy release.
Two hadronic iron/scintillator-strip sampling calorimeters (MUV1,2) and an array of scintillator tiles located behind 80 cm of iron (MUV3,  with 400~ps time resolution)  supplement the pion/muon identification system.
A lead/scintillator shashlik  calorimeter (IRC) located in front of the LKr, covering an annular region between 65 and 135~mm from the Z axis, and a similar detector (SAC) placed on the Z axis at the downstream end of the apparatus ensure the detection of photons down to zero degrees in the forward direction.  Additional counters  (MUV0, HASC) installed at optimized locations provide hermetic coverage for charged particles produced in multi-track kaon decays.

Most detectors are read out  with TDCs, except  LKr and  MUV1,2 read out with 14-bit FADCs, and IRC, SAC read out with both.
With the exception of GTK and STRAW, read out by specific boards, 
the  TDC modules are mounted on custom-made  boards (TEL62),  which both produce trigger information  and perform data read out. 
Calorimeters use a dedicated processor for triggering~\cite{l0calo}.

A low-level trigger  (L0) exploits the logical signals (primitives) produced by the RICH, CHOD, NA48-CHOD, LKr, LAV and MUV3. A dedicated board combines the primitives to build and dispatch trigger decisions to the detectors for data readout~\cite{l0tp}. A software trigger (L1) processes data from KTAG, LAV and STRAW to produce higher level  information 
exploiting reconstruction algorithms similar to those used offline (Sec.~\ref{sec:datareco}),
but adapted to the online environment.

The data sample is obtained from about $5\times10^4$ SPS spills recorded  
during  one month of data-taking in 2016.  
The main trigger chain (called  PNN) is defined as follows.
The L0 trigger requires a signal in RICH to tag a charged particle in coincidence within 10~ns with: a signal in at least one CHOD tile; no signals 
in  opposite CHOD quadrants to reduce \pppc decays; no signals in MUV3 and LAV12 to reduce \mn and \pp decays;  and LKr energy 
deposit below 20~GeV to suppress \pp decays. The L1 trigger requires: a kaon identified in KTAG and signals in at most 
two blocks of each LAV station, within 10~ns of the L0 trigger RICH 
time; at least one STRAW track corresponding to a particle with momentum below 50 GeV/$c$ which forms a vertex with the nominal beam axis upstream of the first STRAW chamber. The 
analysis also uses data taken with a minimum-bias L0 trigger based on NA48-CHOD information downscaled by a factor of 400 (``control triggers") to measure efficiencies and estimate backgrounds. 
Events collected with the PNN (control) trigger are referred to as ``PNN-triggered events'' (``control events'')  in the following.

\section{Reconstruction and calibration}
\label{sec:datareco}
The KTAG channels are time-aligned, and signals are grouped within 2~ns wide  windows to define  candidates. The KTAG candidate time is used as a reference to adjust the time 
response of the other sub-detectors. Signals from the GTK stations grouped within 2~ns of a KTAG candidate form a beam track. Fully reconstructed \pppc decays in the STRAW spectrometer are used to align the
GTK stations transversally to a precision better than 100~$\mu$m and tune the GTK momentum scale.

The STRAW reconstruction relies on the NA48-CHOD time as a reference to determine the drift time. 
Space-points in the chambers describing a path compatible with the magnetic bending define a track, and its parameters are obtained using a Kalman-filter fit.
The $\chi^2$ value and the number of space-points characterize the track quality. Straight tracks collected with the magnet off serve to align the straw tubes to 
30~$\mu$m accuracy. The average value of the \kp mass reconstructed for \pppc decays provides  fine tuning of the momentum scale to  a permille  precision.

Two  algorithms reconstruct  RICH candidates, both grouping signals from PMs in time around the L0 trigger. The first one makes use of a STRAW track as a seed to build a  RICH ring and compute a likelihood for several mass hypotheses ($e^+$, $\mu^+$, $\pi^+$ and $K^+$). 
The second one (``single-ring'') 
fits the signals to a ring assuming that they are produced by a single particle, with the fit $\chi^2$ characterizing  the quality of this hypothesis. 
Positrons are used to calibrate the RICH response and align 
the twenty RICH mirrors  to a precision of $30$~$\mu$rad~\cite{richperf}.

The CHOD candidates  are defined by the response of the two SiPMs reading out  the same tile. Signals in crossing horizontal and vertical 
slabs compatible with the passage of a charged particle form  NA48-CHOD candidates; time offsets depending on the intersection position account for the effect of
light propagation along a slab.

Groups of LKr cells with deposited energy within 100~mm of a seed form LKr candidates (clusters). A seed is defined by a cell in which an energy of at least 250~MeV is released. Cluster energies, 
positions and times are reconstructed taking into account  energy calibration, non-linearity, energy sharing for nearby clusters and noisy cells. 
The final calibration is performed using positrons from \kethree ~decays. 
An additional reconstruction algorithm is applied to maximize the  photon veto efficiency. This 
is achieved by defining candidates as sets of cells with at least 40~MeV energy, closer than 100~mm and in time within 40~ns of each other. 

The reconstruction of MUV1(2) candidates relies on the track impact point. Signals in fewer than 8 (6) nearby scintillator strips are grouped to form a candidate,
the  energy of which is defined as the sum of the energies in the strips, calibrated using weighting factors extracted from dedicated 
simulations and tested on samples of $\pi^+$ and $\mu^+$.

Candidates in MUV3 are defined by time coincidences of the response of the two PMs reading the same tile. The time of a candidate is defined by  the later of the two PM signals, to avoid the 
effect of the time spread induced by the Cherenkov light produced by particles traversing the PM window.

The CHANTI candidates are defined by signals clustered in time and belonging either to adjacent parallel bars or to intersecting orthogonal bars. 
Two threshold settings discriminate the CHANTI, LAV, IRC and SAC TDC signals. Thus up to four time measurements are associated with each signal, corresponding to the leading and 
trailing edge times of the high and low thresholds. The relation between the amplitude of the IRC and SAC pulses provided by the FADC readout and the energy release is determined 
for each channel after baseline subtraction using a sample of \pp decays. 

Signal times measured by GTK, KTAG, CHOD and RICH are further aligned on a spill basis to the L0 RICH time, resulting in 20~ps stability through the whole data sample. 
Early checks on reconstructed data help identifying spills with hardware malfunctioning, which are excluded from the  analysis.

Samples of  \kp decays are produced using a Geant4-based~\cite{geant4} Monte Carlo (MC) simulation of the setup and used in the analysis to validate the detector response, 
compute acceptances and  estimate backgrounds. The simulation includes the modeling of the time development of the signals, the response of the front-end electronic and the effects 
of miscalibration as derived from data. Accidental activity is added in GTK and KTAG assuming 300~MHz  beam intensity, and 
using a pileup beam particle library. 
No accidental activity is simulated in the detectors downstream of the last GTK station. 
Simulated data are subjected to the same reconstruction and calibration steps as described above.

\section{Event selection}
\label{sec:selection}
\begin{figure}[t]
  \begin{center}
    \includegraphics[width=0.496\textwidth]{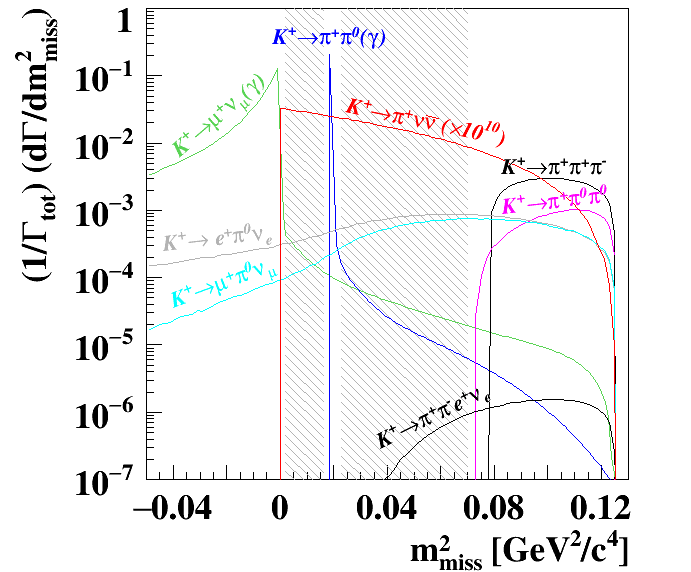}
    \includegraphics[width=0.496\textwidth]{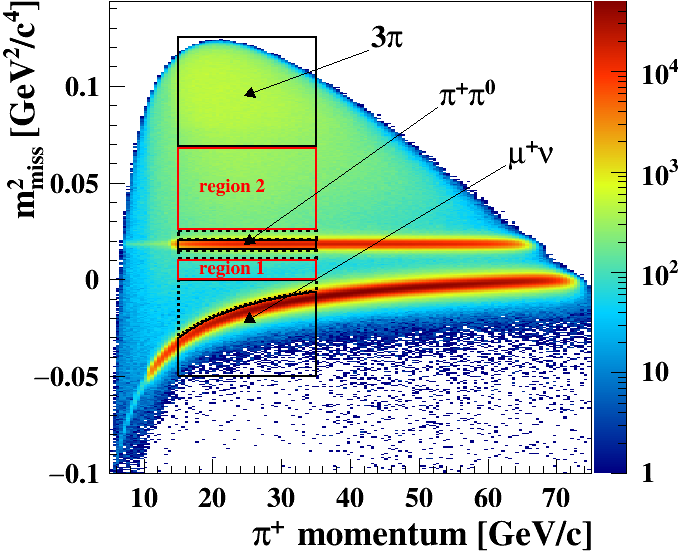}
    \caption{{\bf Left}:  true $\mmis$ distribution of the \pnnc decay and the main \kp decays, computed under \pic mass hypothesis for the charged particle in the final state.
    Signal (red) is multiplied by $10^{10}$ for visibility and the dashed areas show the signal search regions. {\bf {\bf Right}}: reconstructed $\mmis$ as a function of \pic momentum
    for {\it control events} selected without applying \pic identification and photon rejection. Signal regions 1 and 2, as well as the $3\pi$, $\pi^+\pi^0$ 
    and $\mu^+\nu$ background regions are also shown. The control regions located between the signal and background regions are indicated by dashed lines.}
    \label{fig:kinemc}
  \end{center}
\end{figure}
The \pnnc signature consists of a \kp with 4-momentum $p_K$ in the initial state, and a \pic with 4-momentum $p_\pi$ with missing energy in the final state. The 
squared missing mass $\mmis \equiv \left(p_K - p_\pi \right)^2$ is used to discriminate kinematically the main \kp decay modes from the signal. The signal is searched for 
in two $\mmis$ regions on each side of the \pp peak (Fig.~\ref{fig:kinemc}, left). Selection criteria based on $\mmis$ alone are not sufficient to reduce the backgrounds to the desired level, and 
additional suppression by \pic identification and photon rejection is required. The principal selection criteria are listed below.

Up to two positively-charged STRAW tracks are allowed in an event, provided there are no negatively-charged tracks and these tracks do not form a vertex in the FV.  To be accepted, a track must be within the 
RICH, CHOD, LKr, and MUV1,2,3 sensitive regions, and must be associated in position with candidates in the CHOD, NA48-CHOD,  LKr  and RICH. Association in the RICH relies on the consistency 
between the  track direction downstream of the spectrometer magnet and the position of the ring centre. Timing constraints are applied to the LKr and  RICH candidates with 
respect to the NA48-CHOD candidate time. Track times are measured with 100~ps resolution by combining the sub-detector signals associated with each track.

A KTAG candidate with Cherenkov photons detected in at least five out of eight sectors tags a $K^+$. The parent \kp is defined as that  
closest in time to the \pic candidate within 2~ns. The identification of the GTK track of the parent \kp relies on its time coincidence with 
the KTAG and  RICH, and  its geometric compatibility with the STRAW track quantified by the closest distance of  approach (CDA). 
Probability density functions of the time and CDA distributions for both \kp and accidental beam particles  obtained from reconstructed 
\pppc decays  are used to define a discriminant. The GTK track within 0.6~ns of the KTAG time with the highest discriminant value is 
associated with the parent $K^+$.  
An additional requirement is applied on the discriminant  leading to a $K/\pi$  association efficiency value of 75\%, as measured with a
\pppc data sample. The corresponding fraction of  
incorrect $K/\pi$  associations is  below 1\% when the \kp is correctly reconstructed. If the \kp is not reconstructed, the probability of associating a pile-up beam track to the $\pi^+$  is 3.5\%. 

The \kp and \pic tracks define the decay vertex. Charged pions originating from \kp decays upstream of the final 
collimator or from interactions of beam particles in the GTK stations can mimic a \pnnc decay if an accidental beam particle in GTK matches the  $\pi^+$, leading to incorrect vertex reconstruction. 
In addition to $K/\pi$  association, further conditions suppress these events: $Z_\mathrm{vertex}> Z_0$, where $Z_\mathrm{vertex}$ is the vertex longitudinal coordinate and $Z_0$ lies in the range 110--115~m depending on the $\pi^+$ direction; 
the \pic extrapolated to the  final collimator Z plane must lie outside a $200\times1000$~mm$^2$ area centred around the Z axis (``box cut''); no activity in 
CHANTI is allowed within 3~ns of the $\pi^+$;  and fewer than five  GTK tracks may be reconstructed within 2~ns of the KTAG time. To reduce background
from $K^+\rightarrow\pi^+\pi^+\pi^-$ decays and  optimize the $\pi^0$ rejection,  $Z_\mathrm{vertex}$ is required to be upstream of 160--165~m, depending on the track slope.

If two STRAW tracks in the same event satisfy the above conditions, the one closer in time to the trigger and the associated \kp is considered. The CHOD, KTAG, RICH and LKr  candidates
and the GTK track matched to the \pic are required to be the closest to the trigger time. The analysis is restricted to the pion momentum ($P_{\pi^+}$) range of  $(15,35)$~GeV/$c$.
This condition ensures the presence of at least 40 GeV energy in addition to the $\pi^+$, which improves the rejection of backgrounds 
such as $K^+\rightarrow\pi^+\pi^0$. It also enhances the kinematic separation between signal and $K^+\rightarrow\mu^+\nu$  decay, and makes the optimal use of the RICH for $\pi^+ / \mu^+$ separation. 
The $\mmis$ is reconstructed from the \kp and \pic  4-momenta  measured by the GTK and the STRAW. The $\mmis$ resolution at the \pp peak  is about $10^{-3}$~GeV$^2$/$c^4$, which  
determines the choice of two $\mmis$ signal regions 
defined as $(0,0.01)$~GeV$^2$/$c^4$ (``Region~1'') and $(0.026,0.068)$~GeV$^2$/$c^4$ (``Region~2''). Three background regions ($\pi^+\pi^0$, $\mu^+\nu$ and $3\pi$) and suitable control 
regions are also chosen (Fig.~\ref{fig:kinemc}, right). Signal and control regions are kept masked for {\it PNN-triggered events} until the completion of the analysis.  The  $\mmis$ is also computed either assuming  the average \kp beam momentum and direction, or using $p_\pi$ measured by the RICH 
 single-ring algorithm assuming the \pic mass. Conditions imposed on  the $\mmis$ computed in these alternative ways refine the definition of signal regions, reducing the  \pnnc 
acceptance by a further relative 7\% and providing additional background suppression
 in case of mis-reconstruction in STRAW or GTK. 

A multivariate classifier based on 
a Boosted Decision Tree algorithm~\cite{bdt}  
combines 13 variables describing the energy associated with the $\pi^+$, the shape of the clusters 
and  energy sharing between the calorimeters. 
Pion identification with the RICH exploits 
the ratio of likelihoods under the \pic and $\mu^+$ hypotheses. Additional constraints are applied on the particle mass calculated from the ring radius computed  by the single-ring RICH 
algorithm and the momentum measured by the STRAW. The \pic identification efficiency measured in the  $(15,35)$~GeV/$c$ momentum range is 78\% (82\%) with calorimeters (RICH),
and the corresponding probability of \mup mis-identification as \pic is $0.6\times10^{-5}$ ($2.1\times10^{-3}$).
MC simulations reproduce these results with 10--20\% accuracy.
A requirement of no particle detected in MUV3 within 7~ns of the \pic reinforces the MUV3 trigger condition. 

Photon rejection suppresses \pp decays which can mimic the signal if the two-body kinematics is not well reconstructed. The main requirements are: no energy deposited in any LAV station (IRC and SAC)
within 3 (7)~ns of the \pic time; no clusters in the LKr beyond 100~mm from the \pic impact point location within time windows ranging from $\pm5$~ns for 
cluster energies below 5~GeV to $\pm50$~ns above 15 GeV.  Multiplicity rejection criteria
 against photons interacting in the detector material upstream of the LKr  include: no in-time activity in the CHOD and NA48-CHOD 
unrelated to the $\pi^+$ but  in spatial coincidence with an energy deposit of at least 40 MeV  in the LKr; no additional segments reconstructed in the STRAW compatible with the decay vertex; 
no in-time signals in HASC and MUV0; fewer than four extra signals in the NA48-CHOD in time with the $\pi^+$. The resulting  $\pi^0\rightarrow\gamma\gamma$ rejection inefficiency 
is measured  to be $2.5\times10^{-8}$ by counting selected {\it control (PNN-triggered) events} in the $\pi^+\pi^0$ region before (after) photon and multiplicity rejection. Alternatively the single photon detection 
efficiencies of LAV, LKr, IRC and SAC are measured using \pp decays with one  reconstructed photon used to tag  the other photon, and  convolved with a \pp simulation; this  leads to
 a similar  $\pi^0$ rejection estimate. The corresponding signal loss is 34\%, with 10\% (24\%) due to \pic interactions (accidentals). Photon 
and multiplicity rejection is also effective against \pppc and \kefour backgrounds.

\section{Single event sensitivity}
\label{sec:pnnses}
The single event sensitivity is defined as ${\rm SES} =1/(N_K\cdot\varepsilon_{\pi\nu\nu})$, where $N_K$ is the number of \kp decays in the FV and 
$\varepsilon_{\pi\nu\nu}$ is the signal efficiency. The former quantity is computed as $N_K=(N_{\pi\pi}\cdot D)/(A_{\pi\pi}\cdot \text{BR}_{\pi\pi})$, where $N_{\pi\pi}$ is the 
number of \pp decays selected from {\it control events} using the \pnnc criteria except photon and multiplicity rejection, and requiring $\mmis$ to be in the $\pi^+\pi^0$
region; $A_{\pi\pi}=(9.9\pm0.3)$\% is the  acceptance of this selection estimated from MC simulation, where the error is due to the accuracy in the simulation 
of  $\mmis$ resolution; $\text{BR}_{\pi\pi}$ is the \pp branching ratio \cite{pdg} and $D=400$ is the downscaling factor of the control trigger. This leads to $N_K=(1.21\pm0.04_{syst})\times10^{11}$,
where the uncertainty is dominated by  the error on $A_{\pi\pi}$.

The signal efficiency  is evaluated in four 5~GeV/$c$ wide $P_{\pi^+}$ bins as  $\varepsilon_{\pi\nu\nu} =A_{\pi\nu\nu}\cdot\varepsilon_{trig}\cdot\varepsilon_{RV}$.  
Here $A_{\pi\nu\nu}$ is the selection acceptance for the signal;  
$\varepsilon_{trig}$  is the PNN trigger efficiency; 
$1-\varepsilon_{RV}$ is the fraction of signal events discarded by the photon and multiplicity rejection as a consequence of accidental activity in the detectors (Random Veto). 
Other effects inducing signal loss are either included in $A_{\pi\nu\nu}$ or cancel in the ratio to $A_{\pi\pi}$  when 
computing the SES. An example is  the signal loss due to random veto induced by accidental counts in the MUV3 detector.
\begin{figure}[t]
  \begin{center}
    \includegraphics[width=0.496\textwidth]{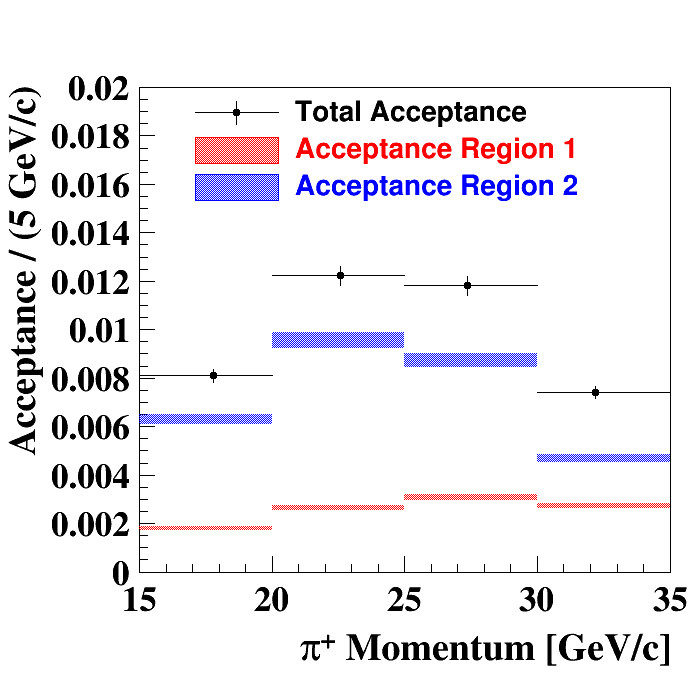}
    \includegraphics[width=0.496\textwidth]{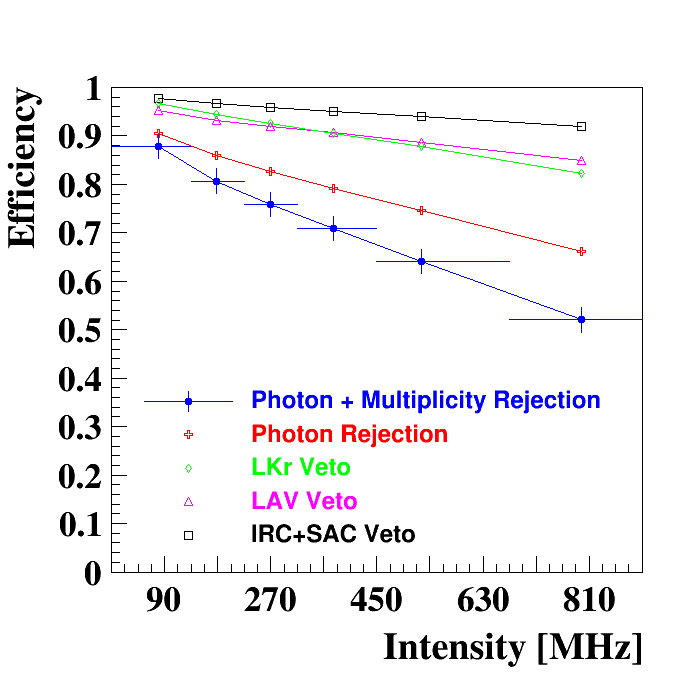}
    \caption{{\bf Left}: Total acceptance $A_{\pi\nu\nu}$ in bins of \pic momentum,  
    and in Regions 1, 2 separately with their respective uncertainties. {\bf Right}: signal efficiency 
    $\varepsilon_{RV}$ in bins of instantaneous beam intensity after photon and multiplicity rejection with total uncertainty, after photon rejection, after IRC and SAC veto only,  
    after LAV veto only, after LKr veto only. Lines are for eye guidance only. The instantaneous intensity (estimated from out-of-time activity in GTK) can vary up to a factor of two within a spill with respect to the average intensity.}
    \label{fig:ses}
  \end{center}
\vspace{-4mm}
\end{figure}

The \pnnc decay is simulated using form factors derived from the \kethree ~decay. The selection acceptance $A_{\pi\nu\nu}=(4.0\pm0.1)$\% is evaluated using MC simulation and includes the particle identification efficiency (Fig.~\ref{fig:ses}, left).  
The main sources of acceptance losses are: detector geometry, $\pi^+$ momentum range, $\mmis$ regions, particle identification and $K^+/\pi^+$ association in the FV.
The uncertainty on $A_{\pi\nu\nu}$ is due to the accuracy of the simulation of the signal losses resulting from photon and multiplicity rejection induced by \pic interactions with the detector material. 

The quantity $\varepsilon_{trig}$ is the product of L0 and L1 trigger efficiencies. The efficiency of each L0  component is measured with {\em control events}. 
These efficiencies, 
with the exception  of the LKr condition, are evaluated with the \pp  sample used for the $N_K$ computation, while  the LKr efficiency is measured using  a \pp sample selected with both photons  in LAV1--LAV11.
The resulting L0 efficiency ranges from 0.93 to 0.85 depending on $P_{\pi^+}$, where the  main losses come from the LKr and MUV3 conditions, with a systematic uncertainty of 0.02.
The L1 trigger efficiency is measured to be $0.97\pm0.01$ using a \pp sample  passing the PNN L0 condition.  The quoted uncertainty reflects the stability during the data taking.

The factor $\varepsilon_{RV}$ is measured as the fraction 
of \mn decays in the data surviving the  photon and multiplicity rejection. The selection used for this measurement 
is similar to the
 \pnnc selection, apart from particle identification (replaced by  positive \mup identification in MUV3 and calorimeters) and the $\mmis$ range requirement. 
The result  is $\varepsilon_{RV}=0.76\pm0.04$;  this quantity is independent of  $P_{\pi^+}$ and depends on beam intensity  as shown in Fig.~\ref{fig:ses}, right. The quoted
value includes a  correction of $+0.02$ based on simulation to account for activity in CHOD and LAV induced by 
$\delta$-rays produced by muons in the RICH mirrors.
The uncertainty on $\varepsilon_{RV}$  is evaluated as the difference between the  loss of acceptance as calculated by simulation and the measured 
$\varepsilon_{RV}$ extrapolated to zero intensity.

The  SES and the corresponding number of SM \pnnc decays expected in the signal regions are:
\begin{eqnarray}\label{eq:signal}
  \mathrm {SES}&= &(3.15\pm0.01_{stat}\pm0.24_{syst})\times10^{-10},\\
  N^{exp}_{\pi\nu\nu}\mathrm {(SM)}&=& 0.267\pm0.001_{stat}\pm0.020_{syst}\pm0.032_{ext}.
\end{eqnarray}
Systematic uncertainties  include those on $N_K$, $A_{\pi\nu\nu}$, $\varepsilon_{trig}$ and $\varepsilon_{RV}$. An additional uncertainty is assigned to beam pileup effects; it is evaluated   
by comparing the acceptances computed including or not the  pileup simulation. 
The external error on $N_{exp}^{\pi\nu\nu}$(SM) comes from the uncertainty of the SM prediction. 

\section{Expected background}
\label{sec:pnnbckg}
Background from \kp decays in the FV is mainly due to $K^+\rightarrow\pi^+\pi^0(\gamma)$, $K^+\rightarrow\mu^+\nu(\gamma)$,~\pppc 
and $K^+\rightarrow \pi^+\pi^- e^+\nu$ decays.
The first three processes may enter the signal regions via $\mmis$ mis-reconstruction.  The estimate of the 
corresponding backgrounds relies on the assumption that $\pi^0$ rejection for $K^+\rightarrow\pi^+\pi^0$, particle identification for \mn and multiplicity rejection 
for \pppc are independent of the $\mmis$ criteria defining the signal regions. Possible violations of this assumption, such as the impact of the radiative component of 
$K^+\rightarrow\pi^+\pi^0(\gamma)$ decay, are investigated separately in a dedicated study. In this framework, the number of expected background events in each signal region (1 and 2)
from these processes is computed as 
$N_{bkg}\cdot f^{kin}$. Here $N_{bkg}$ is the number of {\it PNN-triggered events} remaining in the corresponding background $\mmis$ region after the \pnnc selection; $f^{kin}$ 
(``tails'') is the proportion of background events entering the signal region through tails of $\mmis$ which are modelled separately. The above procedure is applied in four bins 
of \ppic for \pp and $K^+\rightarrow\mu^+\nu$ backgrounds. Background in the control regions is evaluated similarly.
\begin{figure}[t]
\begin{center}
 \includegraphics[width=0.496\textwidth]{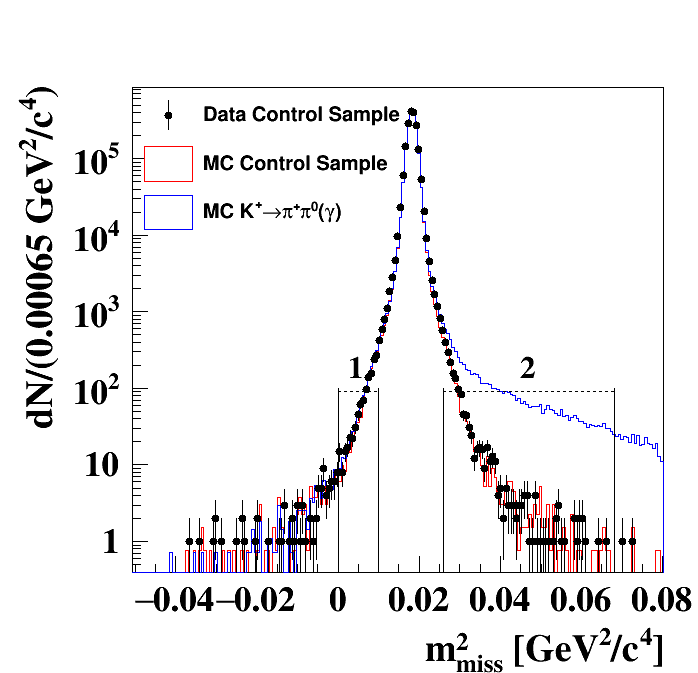}
 \includegraphics[width=0.496\textwidth]{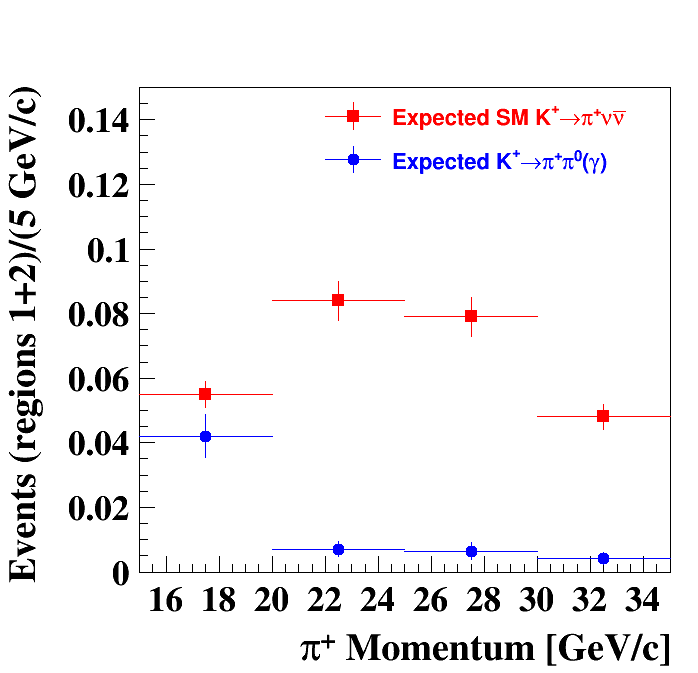}
  \caption{{\bf Left}: reconstructed $\mmis$ distribution of the~\pp {\it control events} selected from data by tagging the $\pi^0$ (dots, see text for details). Two~\pp MC samples are superimposed:
one selected as in data (red line), the other selected as~\pnnc (blue line, referred to as MC $K^+\rightarrow\pi^+\pi^0(\gamma)$ in the legend).  Signal regions 1 and 2 are also shown.
 The MC distributions are normalised to the  data in the $\pi^+\pi^0$ region. {\bf Right}: expected~\ppg background in bins of \pic momentum compared to the expected number  of SM~\pnnc events.}
\label{fig:pp0bckg}
 \end{center}
\vspace{-6mm}
\end{figure}

{\bf {\boldmath$K^+\rightarrow\pi^+\pi^0(\gamma)$} background:} 
forty-two events remain in the $\pi^+\pi^0$ region after the \pnnc selection. The $\mmis$ tails 
in Regions~1 and 2 are at the $10^{-3}$ level and do not  depend on  $P_{\pi^+}$. They are evaluated from  \pp {\it control events} selected with the same criteria as for $N_K$ computation, 
except for the $\mmis$ condition.  
The two photons from the $\pi^0\to\gamma\gamma$ decay are required to be in the LKr acceptance, and 
 \pp decays are reconstructed independently of the measured \pic and $K^+$ tracks by imposing  $\pi^0$ mass, nominal beam momentum and 
missing mass constraints.  Simulation reproduces the tails 
to an accuracy better than 20\%. The kinematic constraints 
do not affect the reconstruction tails in Region~1, but suppress the contribution from \ppgr decays in Region~2 (Fig.~\ref{fig:pp0bckg}, left). 
The estimate of \ppgr (inner bremsstrahlung) 
background is based on simulation \cite{gatti} and single photon detection efficiencies measured from data.
A systematic uncertainty  
is assigned to account for the precision of the above measurement. 
The background dependence on \ppic is shown in Fig.~\ref{fig:pp0bckg},~right and is due to photon detection inefficiency at small angles.
After unblinding the control regions between the $\pi^+\pi^0$ and the two signal regions (Fig.~\ref{fig:kinemc},~right), one event is observed, while $1.46\pm0.16_{stat}\pm0.06_{syst}$ events are expected.
\begin{figure}[t]
  \begin{center}
  \includegraphics[width=0.49\textwidth]{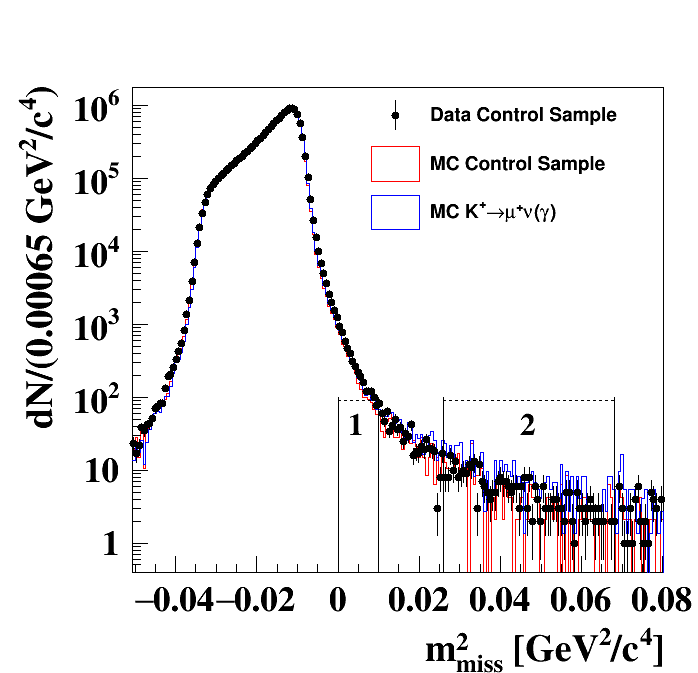}
  \includegraphics[width=0.49\textwidth]{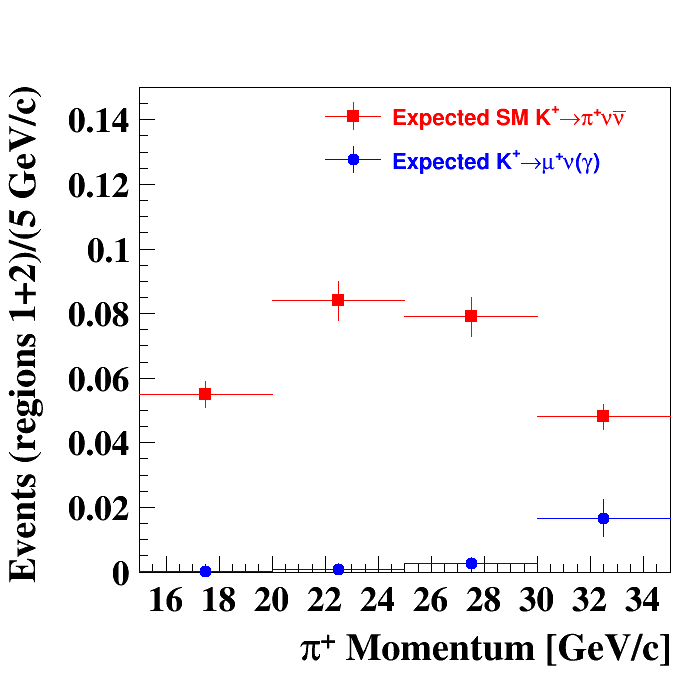}
 \caption{{\bf Left}: reconstructed $\mmis$ distribution of the~\mng control events selected by assigning the $\pi^+$ mass to the $\mu^+$ for data (dots), and for two MC~\mng samples superimposed: 
 one selected as in data (red line), the other selected as~\pnnc without particle identification (blue line, referred as MC~\mng in the legend).  Signal regions 1 and 2 are also shown.
 {\bf Right}: expected~\mng background in bins of \pic momentum compared to the expected number of SM~\pnnc events.}
    \label{fig:km2bckg}
  \end{center}
  \vspace{-6mm}
\end{figure}

{\bf {\boldmath$K^+\rightarrow\mu^+\nu(\gamma)$} background:} forty-five  events remain in the $\mu^+\nu$ region  after the \pnnc selection.
A \mn sample selected as for the $\varepsilon_{RV}$ measurement but without the $\mmis$ condition  is used to evaluate the reconstruction tails (Fig.~\ref{fig:km2bckg}, 
left). Tails in Region~1 range from $5\times10^{-5}$  in the first  $P_{\pi^+}$ bin to $5\times10^{-4}$ in the 
last bin; 
tails in Region~2 are  $2\times10^{-5}$ in all $P_{\pi^+}$ bins. Simulations reproduce these results to an accuracy of 30--40\% in Region~1 and the control region, and better than 
10\% in Region~2. The radiative contribution is included in the measured tails. The calorimetric conditions used to identify \mup in the control sample lead to 
underestimation of the tails in Region~2 by up to 50\%, limited by the current data statistics, and a corresponding systematic uncertainty is assigned to the tail measurement in this region. Mis-measurement 
of \ppic in the STRAW may introduce a correlation between the reconstructed $\mmis$ and \pic identification in the RICH. The effect of this correlation is estimated  
by comparing the RICH performance measured with data \mn events in $\mu^+\nu$ and signal regions. 
This leads to a 30\% uncertainty on the background estimate.
The background dependence on \ppic  
(Fig.~\ref{fig:km2bckg}, right)
is driven by the increase of both the muon mis-identification probability and the tails with $P_{\pi^+}$.
After unblinding the control region, between Region~1 and the $\mu^+\nu$ region (Fig.~\ref{fig:kinemc}, right), two events are observed while $1.02\pm0.16_{stat}\pm0.31_{syst}$ events are expected.

{\bf {\boldmath$K^+\rightarrow\pi^+\pi^+\pi^-$} background:} it populates mainly Region~2 due to  the large $\mmis$. Twenty  events 
remain in the $3\pi$ region after the \pnnc selection. A \pppc sample used to evaluate  the $\mmis$ tails is selected from {\it control events} using a $\pi^+\pi^-$ pair to tag the decay.   
Simulations reproduce the $\mmis$ distribution of this  sample over four orders of magnitude.  
 The tails are conservatively estimated to be  $10^{-4}$, with  a systematic uncertainty of 100\% assigned  to account for  the bias induced by the  above selection, as demonstrated by simulations.
 Multiplicity rejection and kinematic cuts are effective against \pppc decays, 
and the expected background  is almost negligible. 

{\bf {\boldmath$K^+\rightarrow\pi^+\pi^- e^+\nu$} 
background:} it is   characterized by large $\mmis$  and therefore enters Region 2. It is suppressed by its $\cal O$$(10^{-5})$
  branching ratio, multiplicity rejection, particle identification and kinematics. The  approach adopted to estimate the other major backgrounds 
cannot be  used in this case
because the number of particles in the detector acceptance and thus the multiplicity rejection are correlated with the kinematics. The background is therefore estimated from simulation. Out of 6$\times 10^8$ simulated decays, two events pass the \pnnc selection. 
Modifications to the \pnnc selection, such as requiring a $\pi^-$ instead of $\pi^+$, or inverting specific  multiplicity rejection conditions, define exclusive samples with $\cal O$(10)   data events 
passing these new selections. The agreement between the observed and expected numbers of events in these  samples validates the simulation.

{\bf Other backgrounds from {\boldmath$K^+$} decays:} 
the contributions   from \kmthree, \kethree  and $K^+\rightarrow\pi^+\gamma\gamma$ decays are found to be negligible  considering  particle identification 
 and photon rejection performance applied to simulated samples.  Upper limits of  $\cal  O$$(10^{-3})$ events are obtained for these contributions.
\begin{figure}[t]
  \begin{center}
    \includegraphics[width=0.496\textwidth]{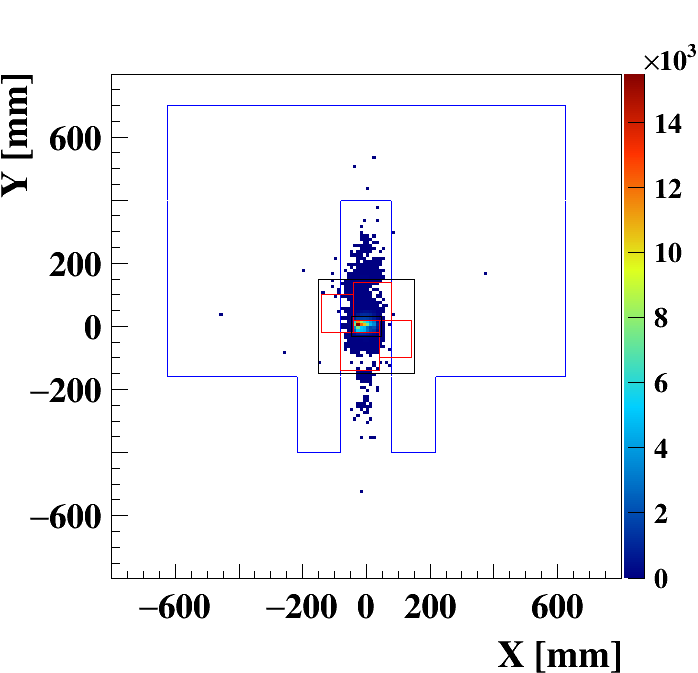}
    \includegraphics[width=0.496\textwidth]{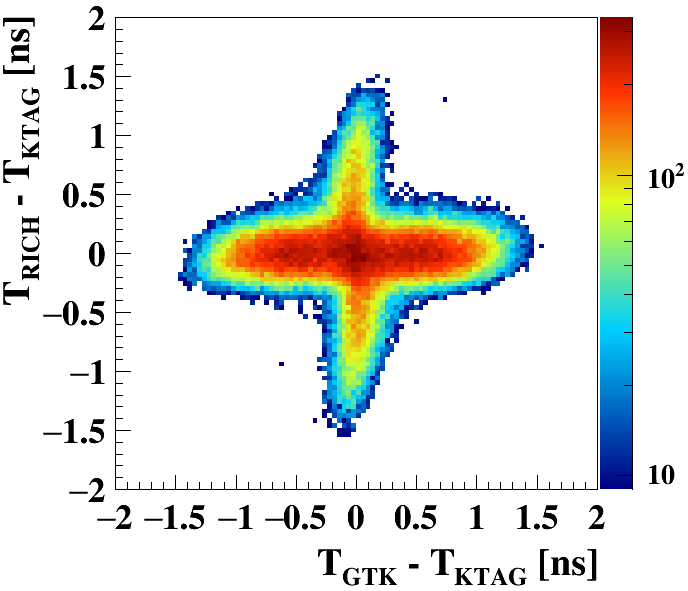}
    \caption{{\bf Left}: transverse position at the FV entrance  of pions from a data sample enriched with upstream events. Blue lines correspond to the contour 
             of the last dipole of the second achromat; red lines show the contour of the final collimator; the black line indicates the acceptance region covered by 
             CHANTI. {\bf Right}: time difference between RICH and KTAG versus that of GTK and KTAG for the same sample of pions.} 
    \label{fig:beambckg}
  \end{center}
  \vspace{-8mm}
\end{figure}

{\bf Upstream backgrounds:}  
they  are due to pions  originating upstream of the FV and are classified as follows.
\begin{itemize}
  \item \pic from \kp decays  between GTK stations 2 and 3, matched to an accidental beam particle;
  \item \pic from interactions of a beam \pic  at GTK stations 2 and 3, matched to an accidental $K^+$;
  \item \pic from interactions of a \kp with material in the beam line, produced either  promptly 
  or as a decay product of a neutral kaon.
\end{itemize}
The interpretation of the upstream events in terms of the above mechanisms is supported by a detailed analysis of a sample selected as \pnnc apart from the conditions defining
the FV and the $K/\pi$  association, which are inverted to enrich the sample with upstream events. 
The  \pic position distribution at the GTK3 Z plane (Fig.~\ref{fig:beambckg}, left) indicates that they originate from upstream decays or interactions in the GTK 
stations and justifies  the $200\times1000$~mm$^2$ box cut (Sec.~\ref{sec:selection}). The distribution of  RICH--KTAG  vs GTK--KTAG  time difference
suggests the accidental origin of these events (Fig.~\ref{fig:beambckg}, right). Installation of an additional shielding insert in 2017 and a 
new collimator of larger transverse size in 2018 have substantially  
reduced the backgrounds  at $|Y| > 100$~mm (Fig.~\ref{fig:beambckg}, left).

 Upstream background estimation is performed with data using a bifurcation technique~\cite{bnl2}. The $K/\pi$ matching ($cut1$) and the box cut ($cut2$) are the
two selection criteria to be inverted. The combination of $cut1$ and $cut2$ defines four samples, denoted  as $A(cut1\cdot cut2)$ corresponding to signal, 
$B(cut1\cdot\overline{cut2})$, $C(\overline{cut1}\cdot cut2)$ and $D(\overline{cut1}\cdot\overline{cut2})$. The number of expected events in sample $A$ is 
$N_A = N_B\cdot N_C/N_D$ assuming $cut1$ and $cut2$ are uncorrelated. This assumption  is validated with data by applying the same procedure to different samples obtained 
modifying the selection conditions. 
The accuracy of the results is limited by the size of the bifurcation samples.

The background  estimates in the signal regions are summarized in Table~\ref{tab:finalbckg}.   
Errors are added in quadrature to obtain the uncertainty on the total expected background.  
\begin{table}[t]
  \begin{center}
   \caption{Summary of the background estimates summed over the two signal regions. 
  } \label{tab:finalbckg}
  \vspace{5pt}
   \begin{tabular}{l|l}
      \toprule
      {Process}                             & {Expected events}              \\\midrule
      $K^+\rightarrow\pi^+\pi^0(\gamma)$  & $0.064\pm0.007_{stat}\pm0.006_{syst}$            \\
      $K^+\rightarrow\mu^+\nu(\gamma)$    & $0.020\pm0.003_{stat}\pm0.006_{syst}$            \\
      $K^+\rightarrow\pi^+\pi^+\pi^-$       & $0.002\pm0.001_{stat}\pm0.002_{syst}$            \\
      $K^+\rightarrow\pi^+\pi^-e^+\nu$      & $0.013^{+0.017}_{-0.012}|_{stat}\pm0.009_{syst}$ \\
      $K^+\rightarrow\pi^0\mu^+\nu$, $K^+\rightarrow\pi^0 e^+\nu$      & $<0.001$            \\
      $K^+\rightarrow\pi^+\gamma\gamma$ & $<0.002$ \\
      Upstream background                   & $0.050^{+0.090}_{-0.030}|_{stat}$                \\\midrule
      Total background                      & ${{0.152^{+0.092}_{-0.033}|_{stat}\pm0.013_{syst}}}$           \\
      \bottomrule
    \end{tabular}
  \end{center}
 \vspace{-5mm}
\end{table}

\section{Result and conclusion}
\begin{figure}[t]
\begin{center}
\begin{minipage}{0.63\linewidth}
\includegraphics[scale=0.17]{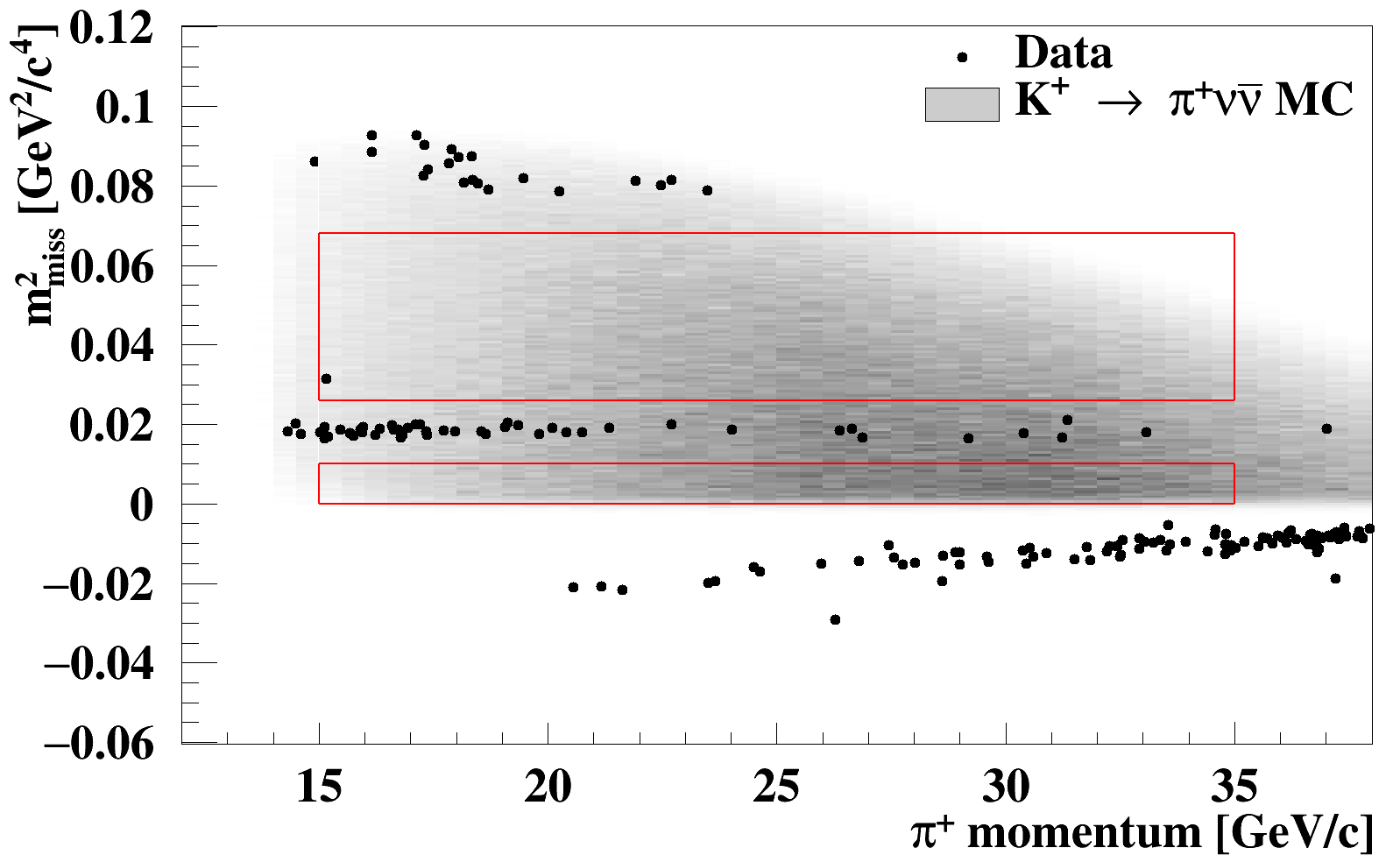}
\end{minipage}
\begin{minipage}{0.34\linewidth}
\hspace{-5mm}
 \includegraphics[scale=0.115]{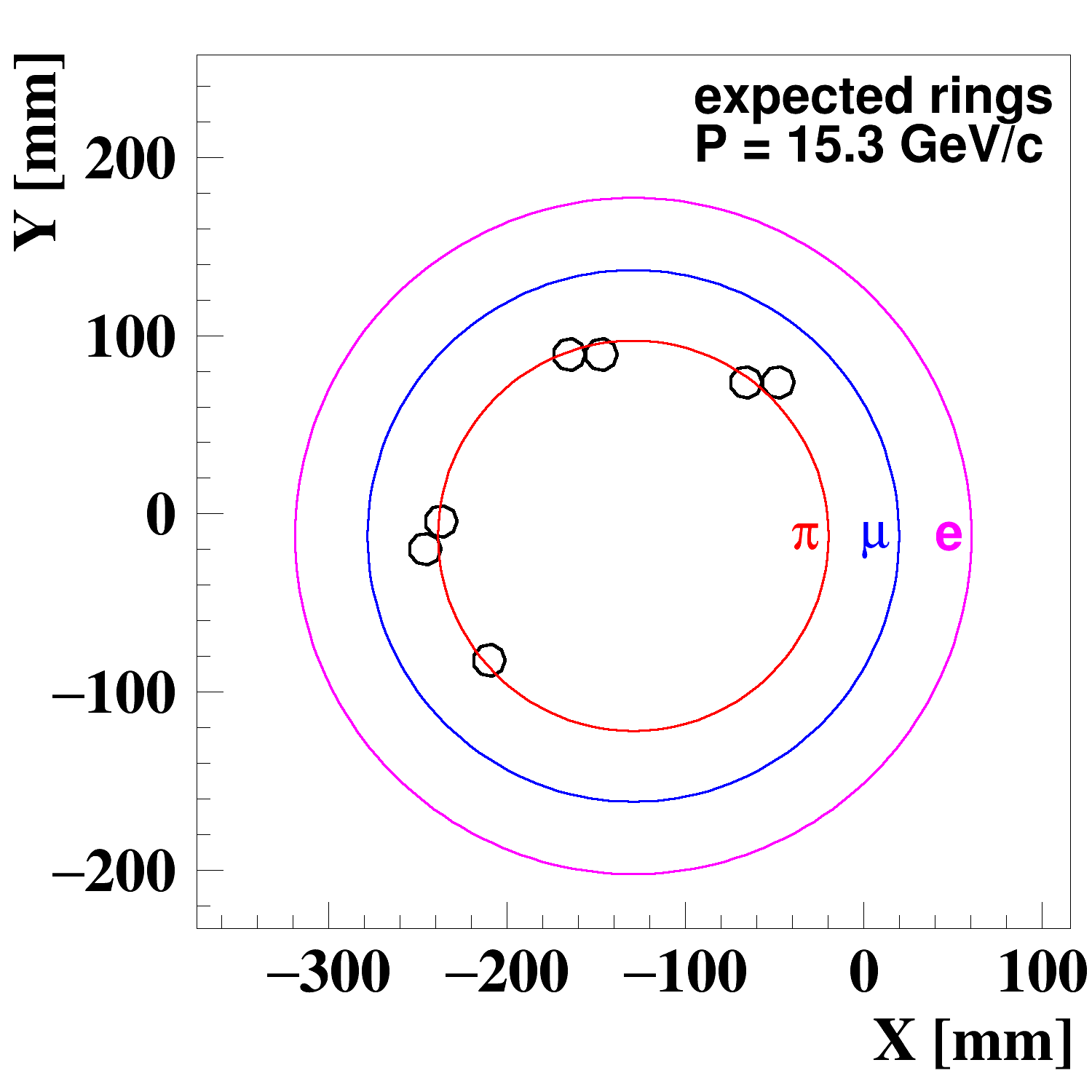}
\end{minipage}
 \caption{{\bf Left}: reconstructed $\mmis$ as a function of \pic momentum for {\it PNN-triggered events} (markers) satisfying the~\pnnc selection, except the  $\mmis$ and \pic momentum criteria.
 The grey area corresponds to the expected distribution of~\pnnc MC events. Red contours define the signal regions. The event observed 
 in Region~2 is shown together with the events found in the control regions. {\bf Right}: Signal in RICH (open circles) detected in the 
 \pnnc candidate event with rings superimposed as built under different particle mass hypotheses.} \label{fig:final}
\end{center}
\vspace{-7mm} 
\end{figure}

After unblinding the signal regions, one event is found in Region~2, as shown in Fig.~\ref{fig:final},~left. The STRAW track 
momentum is 15.3~GeV/$c$. The RICH response is consistent with a \pic hypothesis (Fig.~\ref{fig:final},~right). The track deposits  
50\% (20\%) of its energy in the LKr (MUV1), and  
the $p$-value of the muon hypothesis based on calorimetric identification is 0.2\%.
The candidates  in the RICH, CHOD and LKr associated to the $\pi^+$, as well as the  KTAG and GTK candidates associated to the $K^+$, are mutually consistent in time within one standard deviation of the corresponding resolutions. 
The KTAG candidate is identified by  signals in seven 
of the eight sectors, and the kinematics of the \kp measured with the GTK is consistent with the nominal beam properties. 
The decay vertex properties are $Z_\mathrm{vertex}=146$~m and $\mathrm{CDA} = 1.7$~mm. The $\pi^+$ transverse position extrapolated at the FV entrance is $(x,y)=(-373,30)$~mm. 

The hybrid Bayesian-frequentist approach~\cite{cousin} and the CLs method~\cite{cls} are used for the statistical interpretation 
of the result. A counting experiment analysis is performed considering an expected signal  of 0.267 events
and an expected background  of $0.152 \pm 0.090$ events,  where a symmetric uncertainty is considered.  
The corresponding expected upper limit  
is $\text{BR}_{exp}(K^+\rightarrow\pi^+\nu\bar{\nu})<10\times10^{-10}$ at 95\% CL.
Considering the observation of one event, the $p$-value of the  signal and background hypothesis 
 is 15\%  and the corresponding observed upper limit  is
\begin{equation}\label{eq:br95cl}
  \text{BR}(K^+\rightarrow\pi^+\nu\bar{\nu})<14\times10^{-10}~{\rm at}~95\%~\text{CL}.
\end{equation}

The result, based on 2\% of the total NA62 exposure in  2016--2018, demonstrates the validity of the  decay-in-flight
technique in terms of background rejection and in view of the measurement in progress  using the full data sample.
 
\section*{Acknowledgements}
It is a pleasure to express our appreciation to the staff of the CERN laboratory and the technical
staff of the participating laboratories and universities for their efforts in the operation of the
experiment and data processing.

The cost of the experiment and of its auxiliary systems were supported by the funding agencies of 
the Collaboration Institutes. We are particularly indebted to: 
F.R.S.-FNRS (Fonds de la Recherche Scientifique - FNRS), Belgium;
BMES (Ministry of Education, Youth and Science), Bulgaria;
NSERC (Natural Sciences and Engineering Research Council), Canada;
NRC (National Research Council) contribution to TRIUMF, Canada;
MEYS (Ministry of Education, Youth and Sports),  Czech Republic;
BMBF (Bundesministerium f\"{u}r Bildung und Forschung) contracts 05H12UM5 and 05H15UMCNA, Germany;
INFN  (Istituto Nazionale di Fisica Nucleare),  Italy;
MIUR (Ministero dell'Istruzione, dell'Universit\`a e della Ricerca),  Italy;
CONACyT  (Consejo Nacional de Ciencia y Tecnolog\'{i}a),  Mexico;
IFA (Institute of Atomic Physics),  Romania;
INR-RAS (Institute for Nuclear Research of the Russian Academy of Sciences), Moscow, Russia; 
JINR (Joint Institute for Nuclear Research), Dubna, Russia; 
NRC (National Research Center)  ``Kurchatov Institute'' and MESRF (Ministry of Education and Science of the Russian Federation), Russia; 
MESRS  (Ministry of Education, Science, Research and Sport), Slovakia; 
CERN (European Organization for Nuclear Research), Switzerland; 
STFC (Science and Technology Facilities Council), United Kingdom;
NSF (National Science Foundation) Award Number 1506088,   U.S.A.;
ERC (European Research Council)  ``UniversaLepto" advanced grant 268062, ``KaonLepton" starting grant 336581, Europe.

Individuals have received support from:
Charles University (project GA UK number 404716), Czech Republic;
Ministry of Education, Universities and Research (MIUR  ``Futuro in ricerca 2012''  grant RBFR12JF2Z, Project GAP), Italy;
Russian Foundation for Basic Research  (RFBR grants 18-32-00072, 18-32-00245), Russia; 
the Royal Society  (grants UF100308, UF0758946), United Kingdom;
STFC (Rutherford fellowships ST/J00412X/1, ST/M005798/1), United Kingdom;
ERC (grants 268062,  336581).


\begin{thebibliography}{99}
%
\bibitem{pnntheo1}
A.J.~Buras, D.~Buttazzo, J.~Girrbach-Noe and R.~Knegjens, J. High Energy Phys. 1511 (2015) 33.
%
\bibitem{pnntheo2}
G.~Buchalla and A.J.~Buras, Nucl.\ Phys.\ B 548 (1999) 309.
%
\bibitem{pnntheo3}
A.J.~Buras, M.~Gorbahn, U.~Haisch and U.~Nierste, J. High Energy Phys. 0611 (2006) 002.
%
\bibitem{pnntheo4}
J.~Brod, M.~Gorbahn and E.~Stamou, Phys.\ Review\ D 83 (2011) 034030.
%
\bibitem{pnntheo5}
G.~Isidori, F.~Mescia and C.~Smith, Nucl.\ Phys.\ B 718 (2005), 319.
%
\bibitem{pnnlh}
M.~Blanke, A.J.~Buras and S.~Recksiegel, Eur.\ Phys.\ J.\ C 76 (2016)  182.
%
\bibitem{pnnrs}
M.~Blanke, A.J.~Buras, B.~Duling, K.~Gemmler and S.~Gori, J.\ High\ Energy\ Phys. 0903 (2009) 108.
%
\bibitem{pnnzz}
A.J.~Buras, D.~Buttazzo and R.~Knegjens, J.\ High\ Energy\ Phys. 1511 (2015) 166.
%
\bibitem{pnnsusy1}
G.~Isidori, F.~Mescia, P.~Paradisi, C.~Smith and S.~Trine, J.\ High\ Energy\ Phys. 0608 (2006) 064.
%
\bibitem{pnnsusy2}
T.~Blazek and P.~Matak, Nucl.\ Phys.\ Proc.\ Suppl.\  198 (2010) 216.
%
\bibitem{pnnsusy3}
M.~Tanimoto and K.~Yamamoto, Prog.\ Theor.\ Exp.\ Phys.  2016 (2016) 123B02.
%
\bibitem{pnnlu}
M.~Bordone, D.~Buttazzo, G.~Isidori and J.~Monnard, Eur.\ Phys.\ J.\ C 77 (2017)   618.
%
\bibitem{pnnlq}
C.~Bobeth and A.~J.~Buras, J.\ High\ Energy\ Phys. 1802 (2018) 101.
%
\bibitem{pdg}
M.~Tanabashi  et al.  (Particle Data Group), Phys.\ Rev.\ D 98 (2018) 030001.
%
\bibitem{bnl1}
A.V.~Artamonov et al., Phys.\ Rev.\ Lett.\ 101 (2008) 191802.
%
\bibitem{bnl2}
A.V.~Artamonov et al., Phys.\ Rev.\ D 79 (2008) 092004.
%
\bibitem{na62det}
E.~Cortina~Gil et al., J.\ Instrum.\ 12 (2017) P05025.
%
\bibitem{l0calo}
R.~Aliberti et al., PoS EPS-HEP2017 (2017) 517.
%
\bibitem{l0tp}
D.~Soldi and S.~Chiozzi,  J.\ Instrum.\ 13 (2018) P05004.
%
\bibitem{richperf}
G.~Anzivino et al., J.\ Instrum.\ 13 (2018) P07012.
%
\bibitem{geant4}
J.~Allison et al., Nucl.\ Instrum.\ Methhods A 835 (2016) 186.
%
\bibitem{bdt}
  P.~Speckmayer, A.~Hocker, J.~Stelzer and H.~Voss,
  J.\ Phys.\ Conf.\ Ser.\  219 (2010) 032057.
 %
\bibitem{gatti}
C.~Gatti, Eur.\ Phys.\ J. C 45 (2006) 417.
%
\bibitem{cousin}
R.D. Cousins et al., Nucl.\ Instrum.\ Methods A 595 (2008) 480.
%
\bibitem{cls}
A.L. Read, J.\ Phys.\ G 28 (2002) 2693.
%
\end{thebibliography}
\end{document}